\documentclass[fleqn,usenatbib]{mnras}

\usepackage{newtxtext,newtxmath}
\usepackage[T1]{fontenc}
\usepackage{lineno,hyperref}
\usepackage[american]{babel}
\usepackage[toc,page]{appendix}
\usepackage{amsmath}
\usepackage{graphicx}
\usepackage{subcaption}
\usepackage{listings}
\usepackage{float}
\usepackage{aas_macros}
\usepackage{soul}

\title[The anisotropy of the power spectrum in periodic cosmological simulations]{The anisotropy of the power spectrum in periodic cosmological simulations}

\author[G. R\'acz et al.]{
G\'abor R\'acz,$^{1}$\thanks{E-mail: ragraat@caesar.elte.hu}
Istv\'an Szapudi,$^{2}$
Istv\'an Csabai$^{1}$
L\'aszl\'o Dobos$^{3,1}$
\\
$^{1}$Department of Physics of Complex Systems, ELTE E\"{o}tv\"{o}s Lor\'and University, Pf. 32, H-1518 Budapest, Hungary\\
$^{2}$Institute for Astronomy, University of Hawaii, 2680 Woodlawn Drive, Honolulu, HI, 96822\\
$^{3}$Department of Physics and Astronomy, Johns Hopkins University, 3400 N. Charles Street, Baltimore, MD 21218
}

\date{Accepted XXX. Received YYY; in original form ZZZ}

\pubyear{2020}

\begin{document}
\label{firstpage}
\pagerange{\pageref{firstpage}--\pageref{lastpage}}
\maketitle

\begin{abstract}
    The classical gravitational force on a torus is anisotropic and always lower than Newton's $1/r^2$ law. We demonstrate the effects of periodicity in dark matter only $N$-body simulations of spherical collapse and standard $\Lambda$CDM initial conditions. Periodic boundary conditions cause an overall negative and anisotropic bias in cosmological simulations of cosmic structure formation. The lower amplitude of power spectra of small periodic simulations are a consequence of the missing large scale modes and the equally important smaller periodic forces. The effect is most significant when the largest mildly non-linear scales are comparable to the linear size of the simulation box, as often is the case for high-resolution hydrodynamical simulations. Spherical collapse morphs into a shape similar to an octahedron. The anisotropic growth distorts the large-scale $\Lambda$CDM dark matter structures. We introduce the direction-dependent power spectrum invariant under the octahedral group of the simulation volume and show that the results break spherical symmetry.
\end{abstract}




\section{Introduction}

Cosmological N-body simulations are the principal tools to predict non-linear structure formation at late times. Most implementations use periodic boundary conditions to simulate an infinite universe in finite computer memory. The torus topology has  advantages: the simulation box has a finite volume, its geometry is ideal for three-dimensional Fourier-transforms under translation invariance. Despite the numerical convenience, periodic boundary conditions are not likely to be physical, nor are supported by observations. Indeed, the torus topology runs contrary to the Cosmological Principle that the Universe is isotropic.

We show that on scales comparable to the box size the gravitational force significantly differs from the free-boundary case. The highly anisotropic force is always smaller and affects the evolution of the structures inside cosmological simulations. While earlier work extensively studied the effect of finite simulation volumes \cite{2006MNRAS.370..993B, 2007JApA...28..117P, 2009MNRAS.395..918B}, the effects of the gravitational force modified by periodicity has not been thoroughly investigated. We show that these are significant even at box sizes as large as $L_\mathrm{box} \sim 100h^{-1}$~Mpc.

The simplest way to mitigate the effect of anisotropic gravity is to chose an appropriately large box size where only the linear modes are affected throughout the simulation time. Sometimes this is not feasible due to requirements on mass resolution. Small periodic cosmological simulations are widely used when high resolution is needed, such as in the IllustrisTNG-50 simulation \cite{2019MNRAS.490.3234N}, EAGLE simulations \cite{2015MNRAS.454.2277S} or in the Sherwood simulation suite \cite{2017MNRAS.464..897B}. In this paper, we focus on the power spectrum to quantify the effects of anisotropic gravity induced by the periodic boundary conditions.

\subsection{Gravity in periodic cosmological simulations}

In a smoothed particle simulation of structure formation in an expanding universe, the equations of motion take the form of

\begin{equation}
m_i\ddot{\mathbf{x}}_i = \sum\limits_{j=1; j \neq i}^{N} \frac{m_im_j\mathbf{F}(\mathbf{x}_i-\mathbf{x}_j, h_i, h_j)}{a(t)^{3}} - 2 \cdot m_i \cdot \frac{\dot{a}(t)}{a(t)} \cdot \dot{\mathbf{x}}_i,
\label{eq:Comoving_newtonian}
\end{equation}

where $\mathbf{x}_i$ and $m_i$ are the comoving coordinates and the masses of the particles, $h_i$ and $h_j$ are the softening lengths associated with the particles and $a(t)$ is the cosmological scale factor. The $m_im_j\mathbf{F}(\mathbf{x}_i-\mathbf{x}_j, h_i,h_j)$ vector field is the force law between particles $i$ and $j$, which depends on the softening lengths and the boundary conditions. Since we are interested in the large-scale effect of the force $\mathbf{F}(\mathbf{x}_i-\mathbf{x}_j, h_i,h_j)$, we can neglect the effects of softening by setting $h_i$ softening lengths very small. The Newtonian gravitational field $ \mathbf{F}(\mathbf{x}) = -G \cdot \mathbf{x} \left| \mathbf{x} \right|^{-3} $ of a mass particle centered at the origin, with free boundary conditions, is isotropic, because the magnitude of the force between two point masses depends only on the distance between them. When periodic boundary conditions are considered, Ewald summation \cite{1921AnP...369..253E,1991ApJS...75..231H} should be used and the formula becomes
\begin{equation}
        \mathbf{F}(\mathbf{x}) = \sum\limits_{\mathbf{n}}-G\frac{\mathbf{x}-\mathbf{n}L}{|\mathbf{x}-\mathbf{n}L|^3},
\label{eq:ForcePeriodic}
\end{equation}
where $L$ is the linear size of the periodic box, and $\mathbf{n}=(n_1,n_2,n_3)$ extends over all vectors composed of integer triplets, in theory up to infinity.

\begin{figure*}
    \centering
        \includegraphics[width=0.950\textwidth]{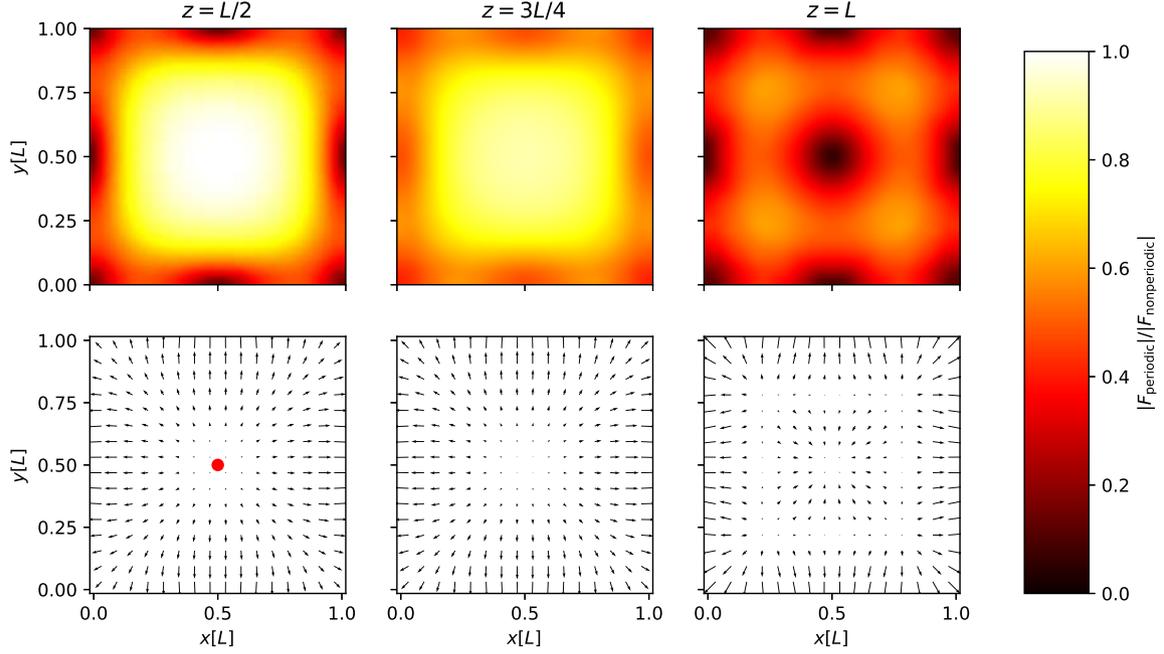}
	\caption{Comparison between the free Newtonian and the periodic gravitational force calculated with the Ewald method. A single particle is placed at coordinates $x=L/2$; $y=L/2$; $z=L/2$ coordinates inside the box with linear size $L$. Each column compares the two force law at different $z$ slice in the simulation volume. \textbf{Top:} The $|\mathbf{F}_\mathrm{periodic}|/|\mathbf{F}_\mathrm{nonperiodic}|$ ratio. \textbf{Bottom:} The $\mathbf{F}_\mathrm{periodic}-\mathbf{F}_\mathrm{nonperiodic}$ difference of the gravitational force fields around the particle. The length of the arrows is proportional to the magnitude of the difference between the forces.}\label{fig:AnisotropicForces}
\end{figure*}

To visualize the consequences of Ewald summation, in Fig~\ref{fig:AnisotropicForces} we plot the difference between the force fields -- with free and periodic boundary conditions -- of a single mass particle placed into the center of simulation box. An obvious consequence of Ewald summation is that the forces are always smaller than in the free Newtonian case, since the multiple images of the point mass act as attractors at a distance. In addition to this, Ewald summation introduces anisotropic distortions to the direction of the force that is significant on the largest scales. For small distances relative to $L$, the force law converges to the isotropic case. When the evolution of the density distribution is followed in Fourier space, it means that, when fluctuations are small enough so that modes evolve independently, all modes but the ones with the largest wavenumbers are affected by anisotropic gravity. In the regime of non-linear structure formation, however, where different modes affect each other, incorrectly treated long wavelength modes can distort smaller-scale fluctuations as well.

\subsection{Symmetries of the simulation cube}
\label{sec:OhSymmetry}

The simulation cube has octahedral symmetry described by the full octahedral group $O_h$ with 48 generators: rotations and reflections.. Every $\mathbf{v}$ vector pointing from the center of the cube towards a specific direction can be transformed into the fundamental tetrahedron by transformations of the $O_h$ symmetry group and the transformed vectors will only occupy $1/48$ of the volume of the cube. This transformation is realized for any $\mathbf{v}=(v_x,v_y,v_z)$ vector by the following algorithm:

\begin{enumerate}
	\item Take the absolute value of each vector component.
	\item Sort them into ascending order.
	\item Normalize the $\mathbf{v}$ vector to $v_z=1$ by dividing by $v_z$. 
\end{enumerate}

\begin{figure}
    \centering
	\begin{subfigure}[b]{0.22\textwidth}
        	\includegraphics[width=\textwidth]{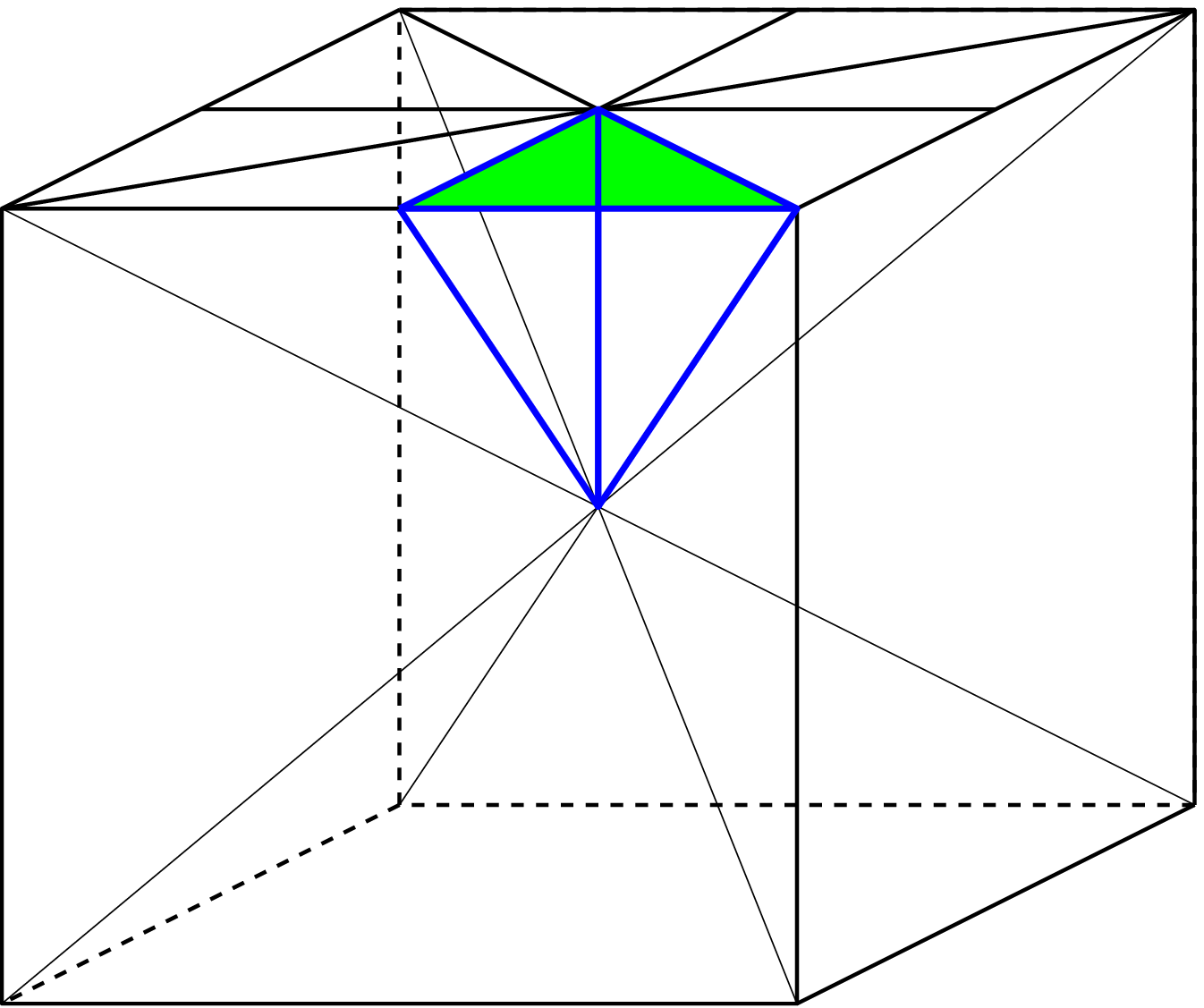}
	\end{subfigure}%
	\hfill
	\begin{subfigure}[b]{0.22\textwidth}
                \includegraphics[width=\textwidth]{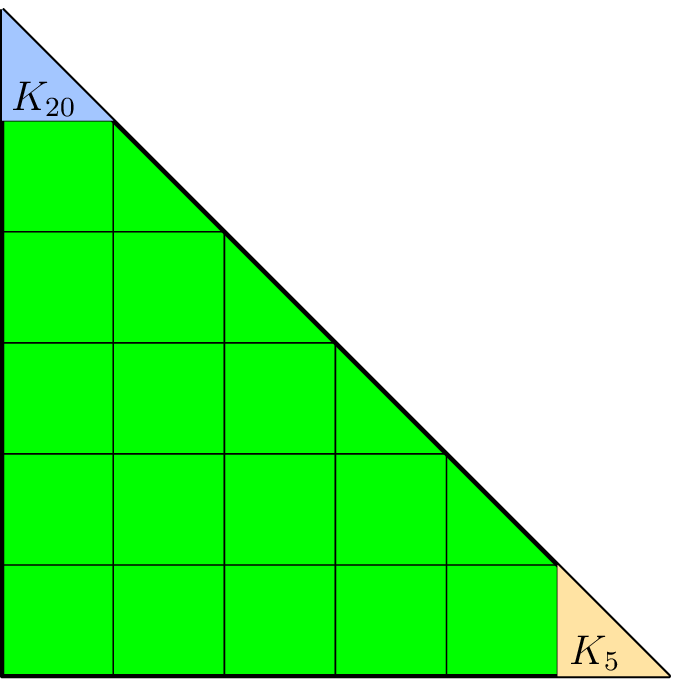}
        \end{subfigure}
	\caption{The fundamental tetrahedra and the fundamental triangles of the cube. \textbf{Left:} the fundamental tetrahedron inside the simulation volume (blue) \textbf{Right:} the fundamental isosceles triangle in the $x-y$ plane, and the 21 disjoint regions $K_i$ we used in the direction-dependent power spectrum calculation. The $\mathbf{\hat{k}}$ vectors pointing towards the regions $K_{5}$, $K_{20}$ are called corner and face directions respectively.}
	\label{fig:OhSymmetry}
\end{figure}

After transformation, every $\mathbf{v}$ vector will point from the origin onto the \textit{fundamental triangle} of the cube. The fundamental triangle is the face of the fundamental tetrahedron that lies on the face of the cube. It is an isosceles triangle that, according to transformation rules defined above, in our case lies in the $x$--$y$ plane, as illustrated in the left panel of Fig.~\ref{fig:OhSymmetry}. There are three notable directions inside the fundamental triangle: the \textit{edge}, the \textit{face} and the \textit{corner} directions which correspond to the vertices of the fundamental triangle.

\subsection{The direction-independent power spectrum}

The Fourier transform of the $\delta(\mathbf{x}) = \rho(\mathbf{x}) / \overline{\rho}-1$ density contrast is defined as
\begin{equation}
        \tilde{\delta}_\mathbf{k} = \frac{1}{V} \int \delta(\mathbf{x})e^{-i\mathbf{k}\mathbf{x}} d^3 x,
\end{equation}
where $V$ is the simulation volume. Since the cosmological principle postulates isotropy, the direction-independent, binned power spectrum
\begin{equation}
        P(k) = \left\langle|\tilde{\delta}_\mathbf{k}|^2\right\rangle_{k-\Delta k/2<|\mathbf{k}|\leq k+\Delta k/2}
        \label{eq:IsotropicPk}
\end{equation}
is often sufficient to describe the statistics of matter density fluctuations. The $\Delta k$ bin size for the spectrum is usually chosen as $\Delta k = 2\pi/L$. The uncertainty of the power spectrum due to sample-variance is
\begin{equation}
    \Delta P(k) = \sqrt{\frac{2}{N_{k}}} \cdot P(k),
    \label{eq:OneSimPkErr}
\end{equation}
where $N_{k}=L^3k^2\Delta k/(2\pi)^2$ is the number of modes per $k$-bin \citep{2016JCAP...04..047S}. 

\subsection{The direction-dependent power spectrum}

The anisotropic nature of forces due to periodic boundary conditions is expected to affect structure formation. To get a statistical picture of this effect, we define the power spectrum that is binned in both wavenumber and direction. Averaging over solid angles (ranges of directions) is necessary to increase the signal. For directional binning, we transform all wave vectors into the fundamental tetrahedron of the simulation box and average the direction-dependent power spectra in each cell of the fundamental triangle, as described in Sec.~\ref{sec:OhSymmetry}.

We divide the fundamental triangle into disjoint partitions denoted by $K_i$, as shown in the right panel of Fig.~\ref{fig:OhSymmetry}. After the transformation of the $\mathbf{k}$ vectors, every vector is assigned into a $K_i$ direction bin. This particular tessellation was chosen to simplify computations and the number of partitions was determined such a way that the resulting, direction-dependent power spectra had sufficiently high signal to noise ratios.

Hence, the direction-dependent power spectrum
\begin{equation}
\mathcal{P}(k,K_i) = \left\langle|\tilde{\delta}_\mathbf{k}|^2\right\rangle_{(k-\Delta k/2<|\mathbf{k}|\leq k+\Delta k/2) \textnormal{and} (\mathbf{\hat{k}} \in K_i)}
        \label{eq:AnsotropicPk}
\end{equation}
is calculated similarly to Eq.~\ref{eq:IsotropicPk}, but the averaging is done only for $\mathbf{k}$ vectors that point towards the $K_i$ direction, as defined in sec.~\ref{sec:OhSymmetry}. The uncertainty of this quantity can be calculated similarly to eq.~\ref{eq:OneSimPkErr} for a single simulation, except now $N_k$ is the number of modes per $k$-bin pointing towards the $K_i$ direction. The is no closed formula for $N_k$ but it can easily be determined algorithmically.

In the following sections, we will overview the effects of periodicity on spherical collapse toy models, and on a series of cosmological dark-matter-only $N$-body simulations.\\

\section{Toy models for periodic structure formation}
In this section, we test the effects of the periodicity and the simulation methods on two distinct toy models for structure formation. We use the standard spherical collapse to test the effects of the periodicity when no compensation is present to shield the tidal fields of the periodic images. On the other hand, the fully compensated Swiss cheese collapse of a spherical region is unaffected by topology due to the Birkhoff theorem, thus it is widely used to assess the accuracy of simulation codes \cite{1995ApJ...452..797C}. In realistic cosmological simulations the effects of the periodicity will fall between these two extremes.

\subsection{Spherical collapse in a periodic box}
\label{sec:SphericalCollapse}

The gravitational collapse of a homogeneous sphere filled with pressureless dust is an important toy model of structure formation. In case of free boundary conditions the sphere contracts isotropically and and analytical solution exist. The solution is parameterized by the angle $\vartheta$, and can be written as
\begin{equation}
\begin{aligned}
	R &= \frac{R_0}{2} \cdot \left( 1+\cos \vartheta \right)\\
	t &= \frac{t_c}{\pi} \cdot \left( \vartheta + \sin \vartheta \right),
\end{aligned}
\end{equation}
where $R$ is the radius of the sphere, $R_0$ is the initial radius, $t$ is the time, and $t_c=\sqrt{3\pi / 32G\rho}$ is the time it takes for the sphere to collapse from rest to infinite density \citep{1999coph.book.....P}.

\begin{figure}
    \centering
        \includegraphics[width=0.50\textwidth]{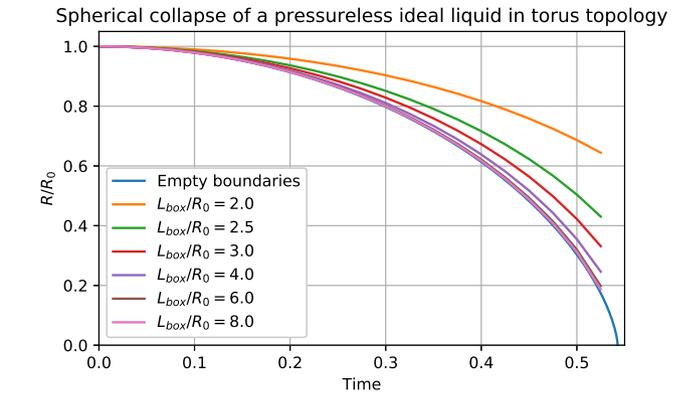}
	\caption{Collapse of a pressureless dust in torus topology. The $R/R_0$ average radius of an initially spherical top-hat overdensity region is plotted as a function of time in different periodic boxes. As the box size increases, the collapse converges towards the free boundary analytic solution.}\label{fig:PeriodicCollapseRR0}
\end{figure}

There is no known analytical solution for spherical collapse in toroidal topology to our knowledge, hence we used the publicly available astrophysical and cosmological simulation code GADGET-2 \cite{2005MNRAS.364.1105S} to demonstrate the effects of periodicity on a self-gravitating dust sphere. To be used as initial conditions, we generated a constant resolution spherical glass with the StePS \cite{2019A&C....2800303R} code and placed the glass at the center of the periodic simulation volume. We set the density inside the sphere to $1$ in internal units and simulated the collapse inside periodic boxes with different linear sizes $L$. The simulations were executed in $G=1$ natural units. Because the collapse is only initially spherical and the anisotropies of the force distort the initial shape, we defined the formula
\begin{equation}
	R(t) = \frac{R_0}{N} \cdot \sum\limits_{i=1}^N \frac{r_i(t)}{r_i(t=0)}
	\label{eq:RforCollapseInTorus}
\end{equation}
to calculate the average radius of the in-falling particle distribution, where $N$ is the particle number, $r_i(t)$ is the distance of the $i$-th particle from the center of the initial sphere at time $t$, and $R_0$ is the initial radius of the sphere. The results of the simulations can be seen in Fig~\ref{fig:PeriodicCollapseRR0}. It is clear, that the spherical collapse is always slower in periodic geometry and it converges towards the free boundary analytic solution as the box size $L$ increases. The same effect can potentially bias the evolution of structure, at least at the largest scales, in cosmological simulations.

\begin{figure}
    \centering
        \includegraphics[width=0.58\textwidth,trim={5mm 0 0 0}]{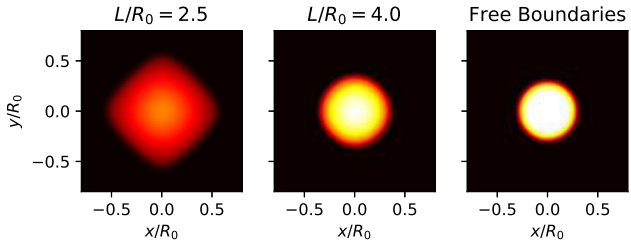}
        \caption{Projected density at $t=0.5$ in spherical collapse simulations with the same initial sphere but different periodic box sizes $L$. The collapse is highly anisotropic when the box size is close to the initial sphere diameter $2R_0$.}\label{fig:PeriodicCollapseDensity}
\end{figure}
The dynamics of the collapse not only differ in speed, but the overall shape of the initial sphere is changing too, as it can be seen in Fig~\ref{fig:PeriodicCollapseDensity}. This anisotropic nature of the spherical collapse in periodic geometry suggests direction dependence or structure formation in periodic simulations.

\subsection{Swiss cheese collapse in a periodic box}

The Swiss cheese spherical collapse is an important test for cosmological simulation codes. The initial condition for these simulations are generated from homogeneous density field. A spherical region inside this simulation with $R_O$ outer radius are selected, and re-scaled to $R_I$ \citep{1993sfu..book.....P, 1995ApJ...452..797C}. The simulation started from this initial density filed will be a spherical collapse inside $R_O$.

The potential outside the outer radius will remain the same during the collapse, as a consequence of the Birkhoff theorem. This ensures that the periodic images of the collapse will not have effect inside a periodic box, so the collapse will be exactly the same as it would be in an infinite volume. Since the geometry of the collapse ensures the isotropy inside the outer radius even in periodic geometry, this fact gives us a chance to test the force calculation in the simulation codes. The only source of anisotropy in these simulations are the errors in the numeric force calculation.

\begin{figure}
    \centering
        \includegraphics[width=0.58\textwidth,trim={5mm 0 0 0}]{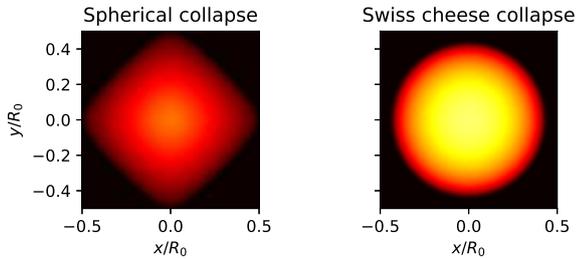}
        \caption{Comparison of the collapsing region in the simulations for the spherical and the Swiss cheese collapse in periodic geometry at $R/R_0=0.43$. See text for details.}\label{fig:SwissCollapse}
\end{figure}

In our test, we used a periodic glass with $5\cdot10^6$ particles as homogeneous particle distribution, and we set the outer diameter of the sphere to match with the linear size of the periodic box. The initial inner radius of the sphere was set to $R_I=L/\sqrt[3]{2}$, so the initial overdensity inside the sphere was $\delta=1$. To have a fair comparison, we have also run a periodic spherical collapse simulation with $R_0=L/\sqrt[3]{2}$ with the same number of inner particles, integration accuracy and the same PM and Tree parameters. We have used the same parameters in our cosmological simulations in the later parts of this paper.

The results can be seen in Fig.~\ref{fig:SwissCollapse}. The collapsed region were spherical through the Swiss cheese collapse, and distorted in the periodic collapse. This eliminates numerical errors as the cause of distortions of the sphere seen in sec.~\ref{sec:SphericalCollapse}.

\section{Periodicity Induced Anisotropy in Cosmological Dark Matter Simulations}

\subsection{Small volume periodic simulations}

A possible definition of a \textit{small} cosmological simulation is that the linear size of simulation box $L$ is comparable to the scale of non-linearity. Since small volume simulations are inherently prone to cosmic variance of the power spectrum \citep{2019ApJ...871..144A}, to be able to see the average effects of periodicity, we run $700$ comoving $\Lambda$CDM $N$-body simulations with five different box sizes. The parameters of the simulations are summarized in Table~\ref{tab:Simulations}, and the cosmological parameters are listed in Table~\ref{tab:SimCosmoParams}. All initial conditions were generated with the 2LPTic code with Planck~2015 cosmological parameters \cite{2016A&A...594A..13P}, and all had different random seeds. The pre-initial conditions were periodic particle glasses generated with GADGET-2.

\begin{table}
	\centering
\begin{tabular}{  l | c c c  }
\hline
	Name & $N_\textnormal{sim}$ & $N_\textnormal{Particles}$ & $L[\textnormal{Mpc}/h]$ \\
\hline
	35Mpch\_2M & $200$ & $2\times10^6$ & 35.0 \\
	50Mpch\_2M & $100$ & $2\times10^6$ & 50.0 \\
	75Mpch\_2M & $100$ & $2\times10^6$ & 75.0 \\
	100Mpch\_2M & $200$ & $2\times10^6$ & 100.0 \\
	250Mpch\_16M & $100$ & $1.6\times10^7$ & 250.0 \\
\hline
\end{tabular}
\caption{The summary of the simulation series.}
\label{tab:Simulations}
\end{table}

\begin{table}
	\centering
\begin{tabular}{  l | c  }
\hline
	Parameter & value \\ \hline
        $\Omega_m$ & $0.3089$ \\
        $\Omega_\Lambda$ & $0.6911$ \\
	$\Omega_k$ & $0.0$ \\
        $H_0 \left[\textnormal{km/s/Mpc}\right]$ & $67.74$\\
        $\sigma_8$ & $0.8159$\\
	$z_{\textnormal{initial}}$ & $63$\\

\hline
\end{tabular}
	\caption{The cosmological parameters common to all simulations. These are based on the Planck 2015 results.}
\label{tab:SimCosmoParams}
\end{table}

\begin{figure}
    \centering
        \includegraphics[width=0.50\textwidth]{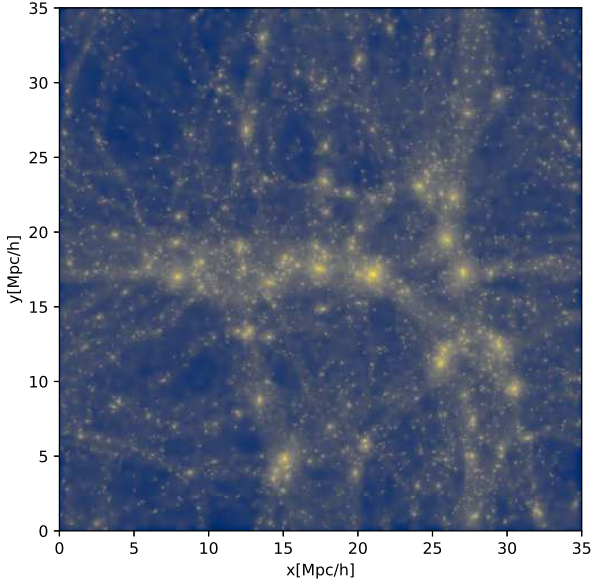}
        \caption{Logarithm of the projected density inside a simulation volume with $L=35\,h^{-1}$\,Mpc. The largest filamentary structures are parallel to the axes of the simulation box at $z=0$. This is a qualitatively typical case of the emerging structures inside a small, periodic volume. See text for detailed description.}\label{fig:LCDMProjectedDensity}
\end{figure}

All simulations were run with GADGET-2. The effects of anisotropic force in the $L=35.0\,h^{-1}$\,Mpc and in the $L=50.0\,h^{-1}$\,Mpc series are striking even for visual inspection of the $z=0$ density fields. Filaments crossing the entire simulation cube are often parallel to the axes and most of the voids resemble cuboids. A typical $z=0$ density field can be seen in Fig.~\ref{fig:LCDMProjectedDensity} as an example. We will quantify these anisotropies in the following sections in terms of the directional power spectra.

\subsection{Large volume reference simulation}

As a comparison to small-volume results, we also run a large, $L=1260.0\,h^{-1}$\,Mpc periodic $\Lambda$CDM simulation with $N=10^{9}$ dark matter particles and the same Planck 2015 parameters. When comparing to small simulations, we made the assumption that the anisotropic effects in the large simulation are negligible in the wavenumber range that overlaps with the longest modes of the small boxes. Consequently, the direction-independent power spectrum of the $L=1260.0\,h^{-1}$\,Mpc simulation was used everywhere as the reference point.

\subsection{Power spectrum bias}

\begin{figure}
    \centering
    \includegraphics[width=0.50\textwidth]{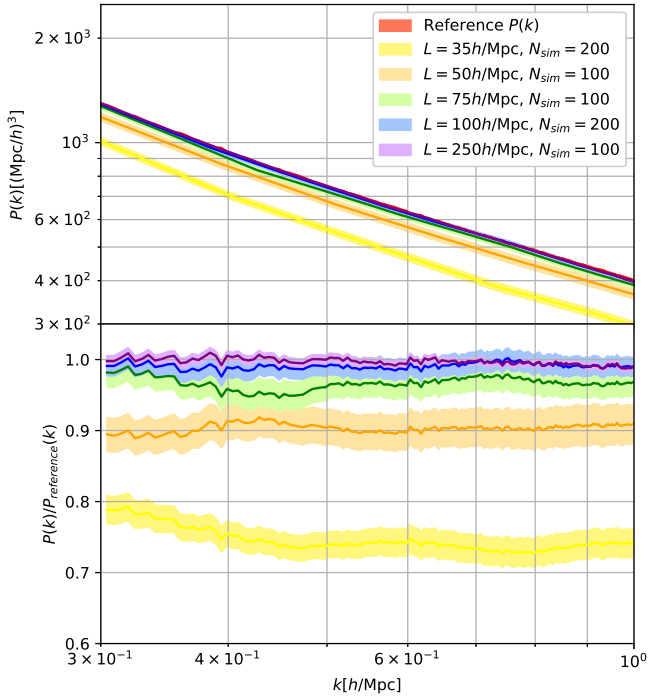}
    \caption{The power spectra of the reference and small box simulations with different periodic box sizes, averaged for many simulations at $z=0$. The bias in the average power spectrum at later times is caused by different growth rates, since the initial spectrum of each simulation was the same.}
    \label{fig:IsotropicPk}
\end{figure}

The results of spherical collapse simulations detailed in Sec.\ref{sec:SphericalCollapse} and plotted in Fig.~\ref{fig:PeriodicCollapseDensity}, suggest that the consistently smaller than Newtonian forces in periodic simulations slow structure formation in smaller boxes. To quantify the expected bias, we computed the average direction-independent power spectrum
\begin{equation}
	\overline{P}(k,L) = \frac{1}{N_{sim}}\sum\limits_{i=1}^{N_{sim}} P_i(k,L)
	\label{eq:AveragePk}
\end{equation}
for every simulation at the $z=0$ final state for each simulation series of the same $L_\mathrm{box}$, where $P_i(k,L)$ is the $z=0$ power spectrum of the simulation with $i$ index and $L$ box size. By taking the
\begin{equation}
	\sigma^2(k,L) = \frac{1}{N_{sim}-1} \sum\limits_{i=1}^{N_{sim}} \left[ P_{i}(k,L) - \overline{P}(k,L) \right]^2
\end{equation}
standard deviation, we estimated the uncertainty of the averaged directional power spectrum as
\begin{equation}
	\Delta \overline{P}(k,L) = \frac{\sigma(k,L)}{\sqrt{N_{sim}}}
	\label{eq:OneSigmaUncertainty}
\end{equation}
for each simulation series of the same $L$. The resulting average direction-independent power spectra and the reference spectrum from the $L=1260.0\,h^{-1}$\,Mpc simulation are plotted in Fig.~\ref{fig:IsotropicPk}. 

\begin{figure}
    \centering
    \includegraphics[width=0.50\textwidth]{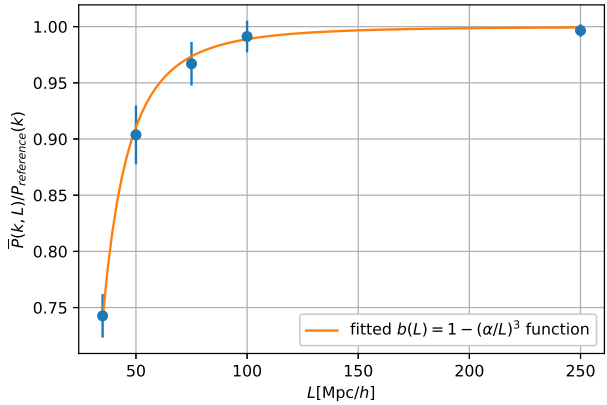}
	\caption{The average bias of the power spectrum with respect to the large reference simulation as a function of the linear box size $L$. The average bias was calculated in the $0.3h\,\mathrm{Mpc}^{-1} < k < 1.0h\,\mathrm{Mpc}^{-1}$ wavenumber range. This bias is a consequence of the missing large-scale modes and smaller forces introduce by periodicity.}\label{fig:IsotropicPkBias}
\end{figure}

We define the power spectrum bias as
\begin{equation}
    B(k, L) = \frac{\overline{P}(k,L)}{P_\mathrm{ref}(k)}.
    \label{eq:Bias}
\end{equation}
To calculate this quantity for the overlapping wavenumber bins of the small box power spectra and the reference power spectrum, one has to interpolate the reference spectrum onto the $k$ grid of the small box spectra. The reference spectrum is sufficiently smooth in the overlapping $k$ ranges so we used cubic spline interpolation. We further average this quantity in the $R_k = \left[ 0.3, 1.0 \right]\,h\,\mathrm{Mpc}^{-1}$ wavenumber range to define a single scalar that represents the average bias of all simulations with a given box size:
\begin{equation}
    B(L) = \frac{1}{N_k} \sum_{k \in R_k} B(k, L).
    \label{eq:Biasavg}
\end{equation}
The wavenumber range was chosen to cover the largest overlapping modes of the small box simulations. Fig~\ref{fig:IsotropicPkBias} shows the average bias of the power spectrum as a function of simulation box size. The initial power spectrum of each simulation was the same, since they were generated by perturbation theory with the same parameters. The measured average power spectrum bias at later times is caused by different growth rates. We found that the small box bias can be fitted as a function of box size in the form of
\begin{equation}
	\overline{P}(k,L) = b(L) \cdot P_\mathrm{ref}(k),
	\label{eq:BiasedPk}
\end{equation}
where
\begin{equation}
	b(L) = 1-\left(\frac{\alpha}{L}\right)^{3}
	\label{eq:Biasfit}
\end{equation}
for box sizes larger than the $\alpha$ value, where $\alpha$ is a free parameter. The best fit value for the parameter is $\alpha = (22.35 \pm 0.02)h^{-1}\,\mathrm{Mpc}$ and the best-fit curve is also plotted in Fig.~\ref{fig:IsotropicPkBias}. This fit is applicable when the box size is considerably larger than the $\alpha$. According to our fit, the anisotropy-induced bias of the direction-independent power spectrum is  less than $1$~per~cent for box sizes $L > 100h^{-1}\,\mathrm{Mpc}$.


One can define the scale of non-linearity by the
\begin{equation}
    \Delta^2(k_\mathrm{nl}) = 1,
\end{equation}
equation, where $\Delta^2(k)=P(k)\cdot k^3/2\pi^2$ is the dimensionless power spectrum. From our large reference simulation, the value of this scale turned out to be $k_\mathrm{nl}=0.2165h\,\mathrm{Mpc}^{-1}$. The corresponding wavelength is $\lambda_\mathrm{nl}=2\pi/k_\mathrm{nl} = 29.02h^{-1}\,\mathrm{Mpc}$, the same order as $\alpha$. This suggests that the observed bias of the small box power spectra is indeed a non-linear effect.

\cite{2016JCAP...04..047S} has found a similar but much larger bias for a single realization of a $L=128\textnormal{Mpc}h^{-1}$ simulation. It is likely that this result is mainly due to cosmic variance.

\begin{figure}
    \centering
    \includegraphics[width=0.50\textwidth]{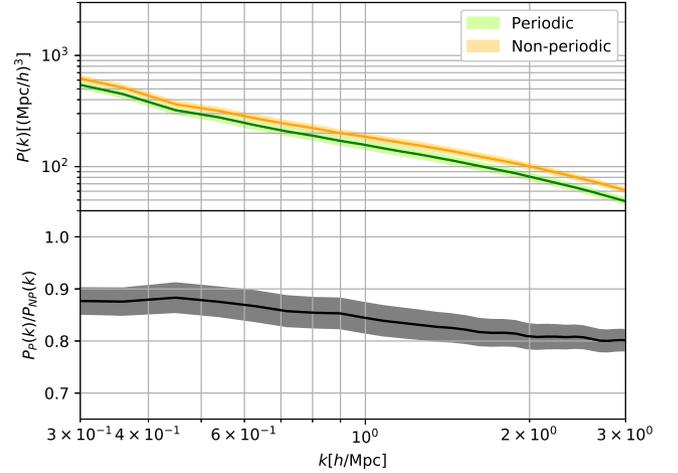}
	\caption{The comparison of the periodic and the non-periodic boundary conditions with a pair of 200 $\Lambda$CDM simulations inside a $28\textnormal{Mpc}h^{-1}$ cubical window. The initial conditions of the two simulation series were the same, and the linear size was $L=35\textnormal{Mpc}h^{-1}$. \textbf{Top:} The power spectrum of the simulations. \textbf{Bottom:} The ratio of the averaged $P_{P}(k)$ periodic and the $P_{NP}(k)$ non-periodic power spectrum.}\label{fig:nonperiodicPk}
\end{figure}

Since the observed power loss is in the non-linear regime, it is plausible that the lack of super-simulation modes, i.e. lack of quasi-linear beat coupling amplification, also contributes to the effect. 
For separating the two effects, we run an ensemble of non-periodic simulations based on our $L=35\textnormal{Mpc}h^{-1}$ boxes shown earlier. The initial conditions were the same, but these were embedded in an infinite homogeneous background universe. We extended the cubic initial conditions to spherical using the original unperturbed glass. We run the simulations with GADGET-2 in comoving coordinates with vacuum boundaries with the same cosmological parameters as the original $35\textnormal{Mpc}h^{-1}$ simulations (type 3 simulation in the GADGET-2 user guide). We have calculated the power spectrum for the new and the original simulations inside a $28\textnormal{Mpc}h^{-1}$ window in the center of the simulations to minimize the boundary effects introduced in the new simulations. The averaged power spectra and their ratio calculated from our 200 periodic and 200 non-periodic simulation can be seen in Fig~\ref{fig:nonperiodicPk}, and shows $14\%$ bias in the $0.3h\,\mathrm{Mpc}^{-1} < k < 1.0h\,\mathrm{Mpc}^{-1}$ wavenumber range. Since the measured bias for the $L=35\textnormal{Mpc}h^{-1}$ boxes was $25\%$, this result suggest that the effect of the missing mode-coupling and periodicity are in the same order of magnitude, and they are equally important in small simulations.
We also re-simulated our $L=35\textnormal{Mpc}h^{-1}$ boxes with quasi-periodic force calculation \citep{1991ApJS...75..231H} that guarantees $1/r^2$ Newtonian gravitational interaction with similar results.

\subsection{The direction-dependent power spectrum}

\begin{figure}
    \centering
        \includegraphics[width=0.50\textwidth]{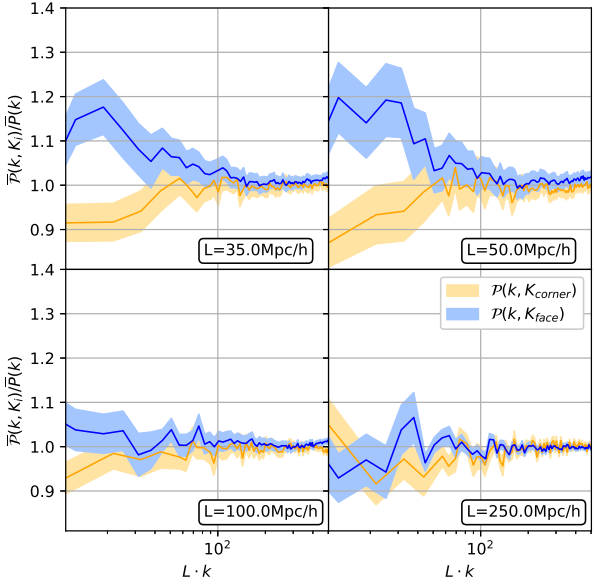}
        \caption{The ratios of the direction-dependent power spectra and the direction-independent $P(k)$ spectrum of the same simulation series, for four different linear box sizes at $z=0$. The spectra are averaged over directions within the $K_5$ and $K_{20}$ regions of the fundamental triangle of the simulation cube, corresponding to the wavenumber vectors pointing towards the corner and face directions.}\label{fig:AnisotropicPk}
\end{figure}

The anisotropic nature of forces in periodic simulations distorts the emerging structures and the rate of structure growth is different in certain, preferred directions of the simulation box, as it can be seen in Fig.~\ref{fig:LCDMProjectedDensity}. To get a statistical picture of this effect, we calculated the direction-dependent power spectrum for every simulation, as introduced in Eq.\ref{eq:AnsotropicPk}. We partitioned the fundamental triangle into 21 disjoint $K_i$ regions, as it can be seen in the right side of fig~\ref{fig:OhSymmetry}, and calculated the 
\begin{equation}
	\overline{\mathcal{P}}(k,K_i,L) = \frac{1}{N_\mathrm{sim}}\sum\limits_{j=1}^{N_\mathrm{sim}} \mathcal{P}_{j}(k,K_i,L)
	\label{eq:AverageAnisotropicPk}
\end{equation}
averaged direction-dependent power spectrum for all of our simulation series of the same box size $L$. We calculated the standard deviation of this quantity as
\begin{equation}
	\sigma^2_{\mathcal{P}}(k,K_i,L) = \frac{1}{N_\mathrm{sim}-1} \sum\limits_{j=1}^{N_\mathrm{sim}} \left[ \mathcal{P}_{j}(k,K_i,L) - \overline{\mathcal{P}}(k,L) \right]^2
\end{equation}
and estimate the uncertainty of the average in form of
\begin{equation}
	\Delta \overline{\mathcal{P}}(k,K_i,L) = \frac{\sigma_{\mathcal{P}}(k,K_i,L)}{\sqrt{N_\mathrm{sim}}}.
	\label{eq:OneSigmaUncertaintyPa}
\end{equation}
We plot the ratio of the averaged direction-dependent power spectra to the direction-independent power spectrum of the same simulation series as a function of $L \cdot k$ in Fig.~\ref{fig:AnisotropicPk}. The blue and the orange curves show the direction-dependent power spectra in the face and corner directions. The former are significantly larger than the direction-independent spectrum if the simulation size is smaller than $100h^{-1}\,\mathrm{Mpc}$.

\begin{figure}
    \centering
        \includegraphics[width=0.50\textwidth]{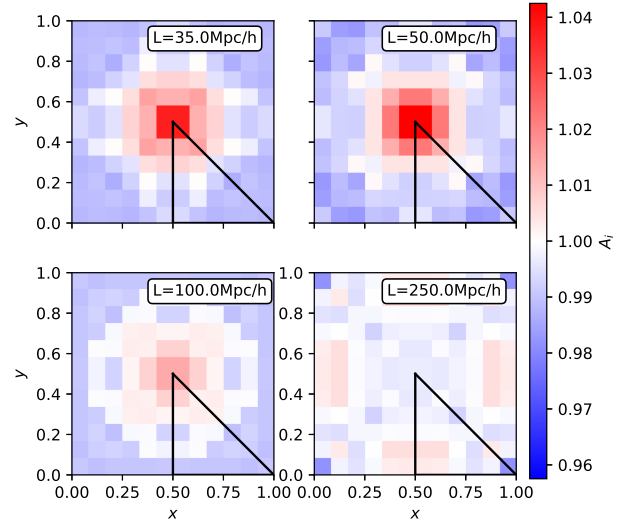}
	\caption{The $A_i(k_\mathrm{min},k_\mathrm{max},L)$ value plotted on a face of the simulation cube for the $L=35h^{-1}\mathrm{Mpc}$, $L=50h^{-1}\mathrm{Mpc}$, $L=100h^{-1}\mathrm{Mpc}$ and $L=250h^{-1}\mathrm{Mpc}$ simulations with $k_{min}=2\sqrt{3}\pi/L$ and $k_{max}=16 \cdot 2\sqrt{3}\pi/L$ limiting wavenumbers. The fundamental triangle (outlined with strong black lines) was divided into 21 disjoint $K_i$ partitions, as it can be seen in Fig~\ref{fig:OhSymmetry} and repeated 8 times to give out the entire face of the simulation cube. The relative power excess towards the face directions are significant for the $L=35h^{-1}\mathrm{Mpc}$ and $L=50h^{-1}\mathrm{Mpc}$ simulations. For larger periodic boxes, the signal is significantly smaller than the noise, as it can be seen in Fig.~\ref{fig:Anisotropy2L}.}
	\label{fig:AiHeatmap}
\end{figure}

We define the signal to noise weighted ratio of the direction dependent and standard power spectrum as
\begin{equation}
\begin{split}
    &A_i(k_\mathrm{min},k_\mathrm{max},L) = \\
    &= \left(\frac{1}{\sum_{\substack{\hat{\mathbf{k}}\in K_i \\ k_{min}<k<k_{max}}} 1/\Delta A(k,K_i,L)}\right) \cdot \\
    & \cdot \sum_{\substack{\hat{\mathbf{k}}\in K_i \\ k_{min}<k<k_{max}}} \frac{1}{\Delta A(k,K_i,L)}\left(\frac{\overline{\mathcal{P}}(k,K_i,L)}{\overline{P}(k,L)}\right)
\end{split}
\label{eq:Anisotropy}
\end{equation}
to measure the anisotropy towards the direction $K_i$ for a given $k_\mathrm{min} < k < k_\mathrm{max}$ wavenumber range, where
\begin{equation}
    \Delta A(k,K_i,L) = \sqrt{\left(\frac{\Delta\overline{\mathcal{P}}(k,K_i,L)}{\overline{\mathcal{P}}(k,K_i,L)}\right)^2+\left(\frac{\Delta\overline{P}(k,L)}{\overline{P}(k,L)}\right)^2} \cdot \frac{\overline{\mathcal{P}}(k,K_i,L)}{\overline{P}(k,L)}
\end{equation}
is the variance of the ratio, and $N_k(k_\mathrm{min},k_\mathrm{max},K_i)$ is the number of discrete $k$ wavenumber vectors between $k_{min}$ and $k_{max}$ pointing towards the $K_i$ fundamental triangle cell. We chose $k_{min}=2\sqrt{3}\pi/L$ and $k_{max}=16 \cdot 2\sqrt{3}\pi/L$ as limits on the wavenumber for every small box simulation. Since the power spectrum is calculated at different discrete $k$ values depending on the $\hat{\mathbf{k}}$ direction, we used cubic spline interpolation on the direction-independent spectrum of the same simulation series to calculate $\overline{P}(k,L)$ at each $k$ value. The resulting $A_i$ values can be seen plotted as a heatmap in Fig.~\ref{fig:AiHeatmap} for the $K_i$ cells of the fundamental triangle. For better understanding the geometry, the plot shows the fundamental triangle repeated eight times to represent the entire face of the simulation box.

\begin{figure}
    \centering
        \includegraphics[width=0.50\textwidth]{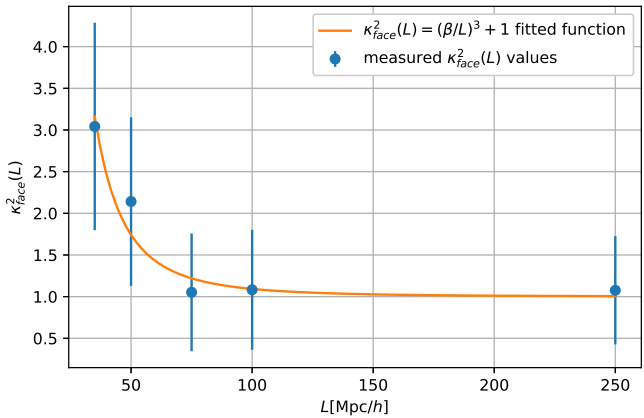}
	\caption{The value of the anisotropy measure $\kappa_\mathrm{face}^2(k_\mathrm{min},k_\mathrm{max},L)$ as a function of the box size $L$, and the fitted function with $k_\mathrm{min}=2\sqrt{3}\pi/L$ and $k_\mathrm{max}=16 \cdot 2\sqrt{3}\pi/L$ arguments. The direction-dependent bias is only significant in our two smallest simulation series.}
	\label{fig:Anisotropy2L}
\end{figure}

In order to take uncertainties of the power spectra into account, in addition to $A_i(k_\mathrm{min},k_\mathrm{max},L)$, we also define
\begin{equation}
 \begin{split}
    &\kappa_i^2(k_\mathrm{min},k_\mathrm{max},L) =\\
    &  = \frac{1}{N_k(k_\mathrm{min},k_\mathrm{max},K_i)}\sum_{\substack{\hat{\mathbf{k}}\in K_i \\ k_\mathrm{min}<k<k_\mathrm{max}}} \frac{(\overline{\mathcal{P}}(k,K_i,L)-\overline{P}(k,L))^2}{\Delta\overline{\mathcal{P}}(k,K_i,L)^2}
 \end{split}
\label{eq:Anisotropy2}
\end{equation}
to characterize the effects of anisotropy in a given $K_i$ direction. Since the power spectrum appears to have the largest bias towards the face directions, we calculated $\kappa_\mathrm{face}^2(k_\mathrm{min},k_\mathrm{max},L)$ values within the wavenumber range $k_{min}=2\sqrt{3}\pi/L$ and $k_{max}=16 \cdot 2\sqrt{3}\pi/L$ range and plotted as a function of $L$ in Fig.~\ref{fig:Anisotropy2L}. Inspired by the fitting formula Eq.~\ref{eq:Biasfit} that we introduced for the isotropic bias, we found that the quantity $\kappa$ can be well fitted in the form of
\begin{equation}
    \kappa_\mathrm{face}^2(k_\mathrm{min},k_\mathrm{max},L) = \left( \frac{\beta}{L} \right)^3 + 1,
\end{equation}
where the free parameter turned out to be $\beta= (45.26\pm2.2)h^{-1}\,\mathrm{Mpc}$. $\beta\simeq 2\alpha$ parameter thus it falls somewhat into the linear regime. According to Fig.~\ref{fig:Anisotropy2L}, the effect of anisotropy is significant at $z=0$, if the simulation box size is smaller than $L \sim 75h^{-1}\,\mathrm{Mpc}$.

\section{Conclusions}

In cosmological $N$-body simulations, the gravitational force is anisotropic and smaller than in the isotropic case, if periodic tidal fields are present. Our spherical collapse simulations illustrate how anisotropic gravity affects the growth of structure: collapse is initially faster in the corner-directions of the toroidal topology.  The structure then morphs into a shape similar to an octahedron, the symmetry underlying the simulation box. After the subsequent non-linear collapse, a star-like configuration emerges from complex anisotropic shell crossing.

With cosmological initial conditions, structure formation is less transparent. Yet, the underlying anisotropy still imprints onto the final structures. Qualitatively, the largest filaments are likely to be orthogonal to the faces in our smallest simulations. The anisotropic structure is due to the direction dependent forces and initial conditions in toroidal topology.

As a first step towards quantifying the effect, we introduced the direction dependent power spectrum invariant under the octahedral group (Eq.~\ref{eq:AnsotropicPk}). The results indeed display a negative bias, and significant distortions in periodic $\Lambda$CDM simulations. There is more power towards the faces then towards the corners, while the overall power is lower. The difference is approximately $15$\% on the largest scales for  $z=0$ in a simulation box of  $L = 35h^{-1}\mathrm{Mpc}$ comoving size, and it becomes sub-percent only above $L > 100h^{-1}\mathrm{Mpc}$. Both the measured bias and the anisotropy scales with the inverse volume of the simulation, with scales $\alpha \simeq 22h^{-1}\mathrm{Mpc}$, and $2\alpha$, respectively. It follows that in simulations smaller than $\alpha$ the effect becomes non-perturbative, therefore it makes no sense to run such small simulations.

We have established that both the negative bias and the anisotropy of the power spectrum are (mildly) non-linear effects. Indeed, linear growth of each mode depends on the Friedman equations which are also solutions in a periodic box with no fluctuations, since periodicity does not matter in that case. Once modes interact, their coupling is affected by the gravitational anisotropy. At the same time, on scales much smaller than the simulation size, gravity approaches the Newtonian $1/r^2$ law. This is why the effect can become significant when the simulation size is comparable to the mildly non-linear scale, $\simeq 30h^{-1}\mathrm{Mpc}$ for $\Lambda$CDM at $z=0$. We verified that there is no significant effect in simulations much larger than this scale. Note that we ran an ensemble of simulations, and small simulations have significant cosmic variance that is larger than the bias we observed. Nevertheless, when such effects are mitigated, the residual bias and anisotropy can be significant. Since the highest-resolution hydrodynamical simulations have small simulation volumes, ray tracing in particular directions can lead to biases. 
Direction-dependent analysis of such simulations is left future work.

In our first study we focused on the power, the most basic statistical measure of large scale structure. By simple extension, it is plausible that if the power spectrum is biased, mass functions would be biased as well. Moreover, our spherical collapse simulation suggest, given the central nature of spherical collapse in perturbative and non-perturbative calculations of non-linear structure formation, that higher order statistics will be affected, including but not limited to their directional dependence. These effects could be investigated by calculating mass functions, stacking structures, such as voids and halos in simulations. Such investigations are left for future work.

The anisotropic effects due to the combination of periodic boundary conditions with small simulation volume can be mitigated by using large enough simulation volumes to derive non-periodic boundary conditions for zoom-in simulations or with isotropic boundary conditions such as the ones used by the StePS \citep{2018MNRAS.477.1949R} simulation method. Alternatively, one can evolve a set of low resolution simulations from a set of random initial conditions, and select one with the most isotropic and unbiased power spectrum and mass function. A similar method were used in the TNG simulation series \citep{2019MNRAS.490.3196P}.

\section*{Acknowledgements}

This work has been supported by the NKFI grants NN 129148 and Quantum Technology National Excellence Program - Project Nr. 2017-1.2.1-NKP-2017-00001. IS acknowledges support from the National Science Foundation (NSF) award 1616974, and from the National Research. LD acknowledges the support of the Schmidt Family Foundation.

\section*{Data availability}

The data underlying this article will be shared on reasonable request to the corresponding author.



\bibliography{AnisotropicPk}

\bsp	
\label{lastpage}
\end{document}